\documentclass[twocolumn, aps, prc, superscriptaddress, floatfix]{revtex4}

\usepackage{multirow}
\usepackage{epsfig}
\usepackage{amsmath}

\usepackage{subfigure}
\usepackage{latexsym}
\usepackage{feynmp}

\usepackage[normalem]{ulem}  % \sout{old text} for strikeout
\usepackage[dvips]{color} % For blue in-text comments and additions

\renewcommand\sout{\bgroup \color{red} \ULdepth=-.5ex \ULset}

\begin{document}

\title{Charm-strange meson production in relativistic heavy ion collisions}

\author{Sungtae Cho}\email{sungtae.cho@kangwon.ac.kr}
\affiliation{Division of Science Education, Kangwon National
University, Chuncheon 24341, Korea}
\author{Su Houng Lee}\email{suhoung@yonsei.ac.kr}
\affiliation{Department of Physics and Institute of Physics and
Applied Physics, Yonsei University, Seoul 03722, Korea}

%\date{\today}
\begin{abstract}

We study charm-strange mesons, or $D_s$, $D_s^*$,
$D_{s0}^*(2317)$, and $D_{s1}(2460)$ mesons by focusing on their
production by coalescence from a quark-gluon plasma in
relativistic heavy ion collisions at $\sqrt{s_{NN}}=5.02$ TeV.
Starting from the investigation of the transverse momentum
distribution of both charm and strange quarks through transverse
momentum distributions of $\phi$ and $D^0$ mesons, we calculate
the transverse momentum distributions and yields of $D_s$,
$D_s^*$, $D_{s0}^*(2317)$, and $D_{s1}(2460)$ mesons based on the
coalescence model. We find that the yield and transverse momentum
distribution of the $D_s$ meson agree well with the experimental
measurements at $\sqrt{s_{NN}}=5.02$ TeV at LHC. We further
evaluate the transverse momentum ratio between $D_s$ and $D^0$
mesons, and investigate the role of light and strange quarks in
the production of charmed mesons in heavy ion collisions. Finally,
we calculate the transverse momentum distribution and yield of the
$D_{s0}^*(2317)$ meson in a four-quark state, and compare to those
of $D_{s0}^*(2317)$ meson in a two-quark state.

\end{abstract}

%\pacs{25.75.-q, 14.65.Dw, 13.60.Rj}

% 14.20.Lq : Charmed baryons
% 13.60.Rj : Baryon production
% 14.65.Dw : Charmed quarks
% 25.75.-q : Relativistic heavy ion collisions

\maketitle

\section{Introduction}

Relativistic heavy ion collision experiments have presented
valuable opportunities to study the production of various kinds of
particles, serving as vital environments for investigating a
system of quantum chromodynamic matter at extreme conditions, the
quark-gluon plasma (QGP) \cite{Adams:2005dq, Adcox:2004mh,
Gyulassy:2004zy}. Enormous energies only achievable in
relativistic heavy ion collisions allow us to study even particles
that are hard to observe in ordinary circumstances
\cite{Thews:2000rj, Andronic:2007bi}. It has been anticipated that
heavy exotic hadrons or multicharmed hadrons, predicted to exist
in nature but have not been discovered yet, may be observed in
relativistic heavy ion collisions \cite{Cho:2010db, Cho:2011ew,
Cho:2017dcy, Cho:2019syk}, presenting us with more various chances
for understanding many aspects of the properties of the QGP.

Since heavy ion collision experiments have been carried out, many
hadrons have been proposed as proper tools to study the QGP, and
strange hadrons are chosen in the early days as those tools to
probe the existence of the QGP, reflecting the strangeness
enhancement observed in relativistic heavy ion collisions than in
elementary $pp$ collisions \cite{Koch:1986ud}. The much heavier
hadron with charm quarks, e.g., the $J/\psi$, has also been
considered to be one of the essential probes in investigating the
properties of the QGP, suggesting the $J/\psi$ suppression due to
color screenings between charm and anti-charm quarks in the QGP
\cite{Matsui:1986dk}.

Moreover, heavy quark hadrons have been regarded as useful in
investigating the properties of the QGP since they can be used to
examine the entire stages in heavy ion collisions. As heavy quarks
are much heavier than light quarks, and their energies are also
higher than any temperatures available in the system, heavy quarks
are produced only at initial hard scatterings in relativistic
heavy ion collisions, propagating through the QGP to be formed as
heavy quark hadrons. With the much more enormous energies
achievable to date, the relativistic heavy ion collision
experiments at $\sqrt{s_{NN}}=5.02$ TeV at the Large Hadron
Collider (LHC) produces a significant number of heavy quarks than
ever before, enabling us to examine the hadronization of those
hadrons from the QGP in more detail.

Among those heavy quark hadrons, charm-strange mesons are special
in the sense that they can play roles in investigating not only
the hadronization of charmed hadrons but also the effects from the
strangeness enhancement compared to non-strange charmed mesons,
i.e., $D^0$ mesons. The $D_s$ meson has been suggested even as a
quantitative tool to study the diffusion and hadronization of
heavy quark hadrons in relativistic heavy ion collisions
\cite{He:2012df}. Hence, recalling the crucial role of the $D_s$
in heavy ion collisions, one may wonder how the $D_s$ meson
reveals its identity in helping to understand the properties of
the QGP when it is produced in systems available at the highest
energy achievable to date. For this reason, we investigate here
the production of $D_s$ mesons, focusing on their transverse
momentum distributions and yields in relativistic heavy ion
collisions at $\sqrt{s_{NN}}=5.02$ TeV at the LHC.

In addition to the $D_s$, we discuss the production of $D_s^*$,
$D_{s0}^*(2317)$, and $D_{s1}(2460)$ mesons as those are major
charm-strange hadrons contributing to the indirect production of
$D_s$ mesons by decay processes \cite{ParticleDataGroup:2024cfk}.
We study the production of the above charm-strange mesons in the
coalescence model, which has been successful in describing many
aspects of the hadron production in relativistic heavy ion
collisions, treating the hadronization as the coalescence process
between constituent quarks in the quark-gluon plasma
\cite{Greco:2003xt, Greco:2003mm, Fries:2003vb, Fries:2003kq}.

Therefore, this work naturally leads us to investigate the
transverse momentum distribution of constituent quarks, i.e.,
light, strange, and charm quarks produced at $\sqrt{s_{NN}}=5.02$
TeV at LHC, as the explicit information on transverse momentum
distributions of constituents at the above energy at LHC is not
currently available. We obtain the transverse momentum
distributions of light, strange, and charm quarks at
$\sqrt{s_{NN}}=5.02$ TeV by referring to the measurements on the
transverse momentum distribution of $\phi$ and $D^0$ mesons.

Then, using the transverse momentum distribution of the $D^0$
meson, we study the transverse momentum distribution ratio between
$D_s$ and $D^0$ mesons. The ratio retains the strange and light
quark contents for the $D_s$ and $D^0$, respectively, after
cancelling the common charm quark, thereby enabling us to
investigate the enhanced production of strange quarks compared to
light quarks in relativistic heavy ion collisions, especially at
low transverse momentum regions.

Here we consider the yield and transverse momentum distribution of
the $D_{s0}^*(2317)$ meson in a two- and a four-quark state in
order to understand the structure of the $D_{s0}^*(2317)$ meson
from the production in relativistic heavy ion collisions. The
structure of the $D_{s0}^*(2317)$ meson remains controversial
since its first discovery by BaBar Collaboration
\cite{BaBar:2003oey} following the confirmation of it by CLEO
Collaboration \cite{CLEO:2003ggt}. The proposed states for the
$D_{s0}^*(2317)$ include a normal charm-strange meson
\cite{Colangelo:2003vg, Bardeen:2003kt}, a $DK$ hadronic molecule
\cite{Barnes:2003dj, Gamermann:2006nm}, and a tetraquark state
\cite{Cheng:2003kg, Bracco:2005kt}.

The production of the $D_{s0}^*(2317)$ meson for the above
suggested states in relativistic heavy ion collisions have already
been discussed, thereby presenting their yields for their
different structures and showing the strong dependence of the
production yield of exotic hadrons including the $D_{s0}^*(2317)$
on their structures \cite{Cho:2010db, Cho:2011ew, Cho:2017dcy}. In
this work, we calculate the transverse momentum distribution as
well as the yield of the $D_{s0}^*(2317)$ for both two- and
four-quark states with new transverse momentum distribution of
charm and strange quarks at $\sqrt{s_{NN}}=5.02$ TeV LHC energy
obtained here. We expect to understand the internal structure of
the $D_{s0}^*(2317)$ meson in more detail from studying the
production of charm-strange mesons in relativistic heavy ion
collisions at $\sqrt{s_{NN}}=5.02$ TeV at LHC.

The paper is organized as follows. In Sec. II, we consider the
production of charm-strange mesons in relativistic heavy ion
collisions, introducing the yield and transverse momentum
distribution of charm-strange mesons within the coalescence model.
We also discuss in Sec. II transverse momentum distributions of
strange and charm quarks by referring to the measurement of
transverse momentum distributions of $\phi$ and $D^0$ mesons by
ALICE Collaboration at LHC. Then, we present in Sec. III yields
and transverse momentum distributions of $D_s$, $D_s^*$,
$D_{s0}^*(2317)$, and $D_{s1}(2460)$ mesons. In addition, we
calculate the yields of the above charm-strange mesons separately
in the thermal model and compare them to the measurements by ALICE
Collaboration at LHC. Furthermore, we discuss the transverse
momentum distribution ratio between $D_s$ and $D_0$ mesons as well
as the transverse momentum distribution of the $D_{s0}^*(2317)$ in
a four-quark state in Sec. III. The Sec. IV is devoted to
conclusions.

\section{Production of charm-strange mesons}

\subsection{Coalescence model for charm-strange meson production}

We consider here the production of charm-strange mesons from charm
and anti-strange quarks in the quark-gluon plasma by coalescence
at the quark-hadron phase boundary in heavy ion collisions.
Adopting the coalescence picture for the formation of the meson
\cite{Fries:2003kq, Fries:2003vb, Greco:2003xt, Greco:2003mm} we
evaluate the yield and transverse momentum distribution of $D_s$,
$D_s^*$, $D_{s0}^*(2317)$, and $D_{s1}(2460)$ mesons.

The yield of the charm-strange meson produced from one charm,
denoted by the subscript $c$ and one anti-strange quark, denoted
by the subscript $\bar{s}$ is given by,
\begin{eqnarray}
&& N_M=g_M\int p_c\cdot d\sigma_c p_{\bar{s}} \cdot
d\sigma_{\bar{s}} \frac{d^3\vec p_c}{(2\pi)^3 E_c}\frac{d^3\vec
p_{\bar{s}}}{(2\pi)^3 E_{\bar{s}}}
\nonumber \\
&& \qquad\times  f_c(r_c, p_c) f_{\bar{s}}(r_{\bar{s}},
p_{\bar{s}}) W_M(r_c, r_{\bar{s}}; p_c, p_{\bar{s}}),
\label{CoalGencharmstrange}
\end{eqnarray}
with $d\sigma_c$ and $d\sigma_{\bar{s}}$ being the space-like
hypersurface element for a charm and an anti-strange quark,
respectively. The subscript $M$ in Eq. (\ref{CoalGencharmstrange})
stands for the kind of charm-strange mesons: $D_s$, $D_s^*$,
$D_{s0}^*(2317)$, and $D_{s1}(2460)$ mesons. $f_q(r_q, p_q)$ is a
covariant distribution function of a quark $q$ satisfying the
normalization condition $\int p_q\cdot d\sigma_q d^3\vec
p_q/((2\pi)^3E)f_q(r_q, p_q)=N_q$, the number of quarks $q$ in the
system. The factor $g_M$ takes into account the possibility of
forming the charm-strange meson from constituent quarks, e.g.,
$g_{D_s}=1/(2\cdot 3)^3$. In the non-relativistic limit, Eq.
(\ref{CoalGencharmstrange}) is reduced to \cite{Greco:2003mm,
Greco:2003xt, Oh:2009zj}
\begin{eqnarray}
&& \frac{d^2N_M}{d^2\vec p_T}=\frac{g_M}{V} \int d^3\vec r d^2\vec
p_{\bar{s}T}d^2\vec p_{cT} \delta^{(2)}(\vec p_T-\vec p_{\bar{s}T}
-\vec p_{cT}) \nonumber \\
&& \qquad\quad\times\frac{d^2N_{\bar{s}}}{d^2 \vec p_{\bar{s}T}}
\frac{d^2N_{c}} {d^2\vec p_{cT}} W_M(\vec r, \vec k),
\label{CoalTransCharmstrange}
\end{eqnarray}
with the assumption of the Bjorken correlation between spatial,
$\eta$ and momentum $y$ rapidities, or the boost-invariant
longitudinal momentum distributions for quarks satisfying
$\eta=y$. In Eq. (\ref{CoalTransCharmstrange}) $\vec r$ and $\vec
k$ are relative distances and transverse momenta between charm and
strange quarks, respectively, in the rest frame of the
charm-strange meson. $\vec k$ contains the information on
different frames through the Lorentz transformation of the
transverse momenta of quarks in the fireball frame, $\vec
p_{\bar{s}T}$ and $\vec p_{cT}$, to those of quarks in the rest
frame of the meson, $\vec p_{\bar{s}T}'$ and $\vec p_{cT}'$
\cite{Scheibl:1998tk, Oh:2009zj}. We consider here the following
quark configurations and relative transverse momenta,
\begin{eqnarray}
&& \vec R=\frac{m_{\bar{s}}\vec r_{\bar{s}}+m_c\vec r_c}
{m_{\bar{s}}+m_c}, \qquad \vec r=\vec r_{\bar{s}}-\vec r_c, \nonumber \\
&& \vec K=\vec p_{\bar{s}T}'+\vec p_{cT}', \qquad \vec
k=\frac{m_c\vec p_{\bar{s}T}'-m_{\bar{s}}\vec
p_{cT}'}{m_{\bar{s}}+m_c}, \label{rel_coordinates}
\end{eqnarray}
with the reduced mass, $\mu=m_sm_c/(m_s+m_c)$. We adopt the
$s$-wave Wigner functions for $D_s$ and $D_s^*$ mesons, and
$p$-wave Wigner functions for $D_{s0}^*(2317)$ and $D_{s1}(2460)$
mesons constructed from harmonic oscillator wave functions,
\begin{eqnarray}
&& W_s(\vec r, \vec k) = 8 e^{-\frac{r^2}{\sigma^2}-k^2
\sigma^2} \nonumber \\
&& W_p(\vec r, \vec k) = \bigg( \frac{16}{3} \frac{r^2}{\sigma^2}
-8+\frac{16}{3} \sigma^2 k^2 \bigg) e^{-\frac{r^2}{\sigma^2}-k^2
\sigma^2} \label{Wigsp}
\end{eqnarray}
with $\sigma^2=1/(\mu\omega)$, where $\omega$ is the oscillator
frequency of the harmonic oscillator wave function. As the
integration of the Wigner function over coordinate spaces, $\vec
r$ give the absolute value square of the wave function in momentum
space \cite{Hillery:1983ms}, the integration of the above Wigner
functions in Eq. (\ref{Wigsp}) leads to \cite{Cho:2014xha},
\begin{eqnarray}
&& \int d^3\vec r W(\vec r, \vec k)=|\tilde{\psi}(\vec
k)|^2  \nonumber \\
&& =\left\{ \begin{array}{ll}
(2\sqrt{\pi}\sigma)^3 e^{-k^2\sigma^2} &:s\textrm{-wave}   \\
\frac{2}{3}(2\sqrt{\pi}\sigma)^3 e^{-k^2\sigma^2}\sigma^2k^2
&:p\textrm{-wave}  \\
\end{array} \right. \label{WigIntcoord}
\end{eqnarray}
Thus, carrying our the integration over the coordinate spaces in
Eq. (\ref{CoalTransCharmstrange}) using Eq. (\ref{WigIntcoord})
results in,
\begin{eqnarray}
&& \frac{d^2N_M}{d^2\vec p_T}=\frac{g_M}{V}(2\sqrt{\pi}\sigma)^3
\int d^2\vec p_{\bar{s}T}d^2\vec p_{cT}e^{-\sigma^2 k^2} \nonumber \\
&& \qquad\quad \times\delta^{(2)}(\vec p_T-\vec p_{\bar{s}T}-\vec
p_{cT})\frac{d^2N_s}{d^2 \vec p_{\bar{s}T}} \frac{d^2N_c}{d^2\vec
p_{cT}}, \label{CoalTransCSs}
\end{eqnarray}
for $D_s$ and $D_s^*$ mesons, and
\begin{eqnarray}
&& \frac{d^2N_M}{d^2\vec p_T}=\frac{g_M}{V}(2\sqrt{\pi}\sigma)^3
\int d^2\vec p_{\bar{s}T}d^2\vec p_{cT}\frac{2}{3}
\sigma^2k^2 e^{-\sigma^2 k^2} \nonumber \\
&& \qquad\quad \times\delta^{(2)}(\vec p_T-\vec p_{\bar{s}T}-\vec
p_{cT})\frac{d^2N_{\bar{s}}}{d^2 \vec p_{\bar{s}T}}
\frac{d^2N_c}{d^2\vec p_{cT}}, \label{CoalTransCSp}
\end{eqnarray}
for $D_{s0}^*(2317)$ and $D_{s1}(2460)$ mesons.

Similarly, we obtain
\begin{eqnarray}
&& \frac{d^2N_\phi}{d^2\vec
p_T}=\frac{g_\phi}{V}(2\sqrt{\pi}\sigma_\phi)^3
\int d^2\vec p_{sT}d^2\vec p_{\bar{s}T}e^{-\sigma_\phi^2 k^2} \nonumber \\
&& \qquad\quad \times\delta^{(2)}(\vec p_T-\vec p_{sT}-\vec
p_{\bar{s}T})\frac{d^2N_s}{d^2 \vec p_{sT}}
\frac{d^2N_{\bar{s}}}{d^2\vec p_{\bar{s}T}}, \label{CoalTransphi}
\end{eqnarray}
for the $\phi$ meson with $g_\phi$ and $\sigma_{\phi}$, being the
degeneracy of the $\phi$, and the size of the $\phi$ meson,
respectively.

\subsection{Strange quark transverse momentum distribution}

As the characteristics of hadrons produced from the quark-gluon
plasma at the quark-hadron phase boundary are inherited from those
of their constituents when they are produced by coalescence, it is
essential to know exact information on quarks in the quark-gluon
plasma as well as the properties of the hadrons in order to
understand the production of hadrons. Here, it is required to
understand the dynamical properties, or the transverse momentum
distribution of strange and charm quarks to evaluate the
transverse momentum distributions or yields of charm-strange
mesons.

Moreover, to explain or predict the properties of hadrons produced
at the LHC $\sqrt{s_{NN}}=5.02$ TeV energy, the information on
constituents at the same energy should be available. However, the
transverse momentum distributions of charm and strange quarks at
$\sqrt{s_{NN}}=5.02$ TeV are not currently available. Therefore,
it is necessary to find the information on strange and charm
quarks first before evaluating the production of charm-strange
hadrons, and one can obtain those from the experimental
measurement of hadrons containing charm or strange quarks.

At first, the $\phi$ meson is chosen for extracting the
information on strange quarks at the LHC $\sqrt{s_{NN}}=5.02$ TeV
energy since it is composed of purely strange quarks, and also has
been already measured at $\sqrt{s_{NN}}=5.02$ TeV
\cite{ALICE:2019xyr}. The $\phi$ mason has a longer lifetime than
the assumed lifetime of the hadronic stage in heavy ion
collisions, and moreover, is hardly affected by the interaction
between light hadrons during the hadronic stage in heavy ion
collisions, such that it is often compared to the $K^*$ meson of
which yield is measured less than that expected by the statistical
hadronization model due to both its short lifetime and
interactions between light hadrons in the hadronic medium
\cite{ALICE:2021ptz, Cho:2015qca}. Therefore, the $\phi$ meson is
considered to keep the information of the strange quarks at the
moment of its production throughout the heavy ion collision
experiments.

The transverse momentum distribution of strange quarks can be
divided into two parts; one from thermal quarks in equilibrium in
low transverse momentum regions, and the other from those quarks
produced from energetic partons in high transverse momentum
regions. The transverse momentum distribution of strange quarks at
low $p_T$ and mid-rapidity is given by \cite{Oh:2009zj,
Cho:2019syk},
\begin{equation}
\frac{d^2N_s}{d^2 \vec p_{sT}}= g_s\frac{V}{(2\pi)^3}m_T
e^{-m_T/T_{eff}}, \label{d2NsdpT2low}
\end{equation}
with the degeneracy factor for color and spin, $g_s$ of strange
quarks, and the transverse mass $m_T=\sqrt{p_{sT}^2+m^2}$, $V$ and
$T_{eff}$ in Eq. (\ref{d2NsdpT2low}) are the coalescence volume
and the effective temperature, respectively. The collective flow
effects are taken into account in the effective temperature, which
is higher than the phase transition temperature expected by the
statistical hadronization model. The transverse momentum
distribution of strange quarks, Eq. (\ref{d2NsdpT2low}) must be
valid in limited $p_T$ ranges, up to $p_{sT}'$, which will be
determined later. The transverse momentum distribution of strange
quarks in the region with their momenta higher than $p_{sT}'$ at
mid-rapidity is adopted from that of charm quarks
\cite{Plumari:2017ntm},
\begin{equation}
\frac{d^2N_s}{d^2 \vec p_{sT}}=s_1
e^{(-s_2(\frac{p_{sT}}{p_{0T}})^{s_3})}
+\frac{s_4}{(1.0+(\frac{p_{sT}}{p_{0T}})^{s_5})^{s_6}}
\label{d2NsdpT2high}
\end{equation}
with parameters, $s_1, s_2, s_3, s_4, s_5$ and $s_6$, having a
power law type transverse momentum distribution of strange quarks
at high transverse momentum regions, and thus enabling to explain
the behavior of the $\phi$ meson transverse momentum distribution
at high transverse momenta. Thus, the full transverse momentum
distribution of strange quarks in the quark-gluon plasma is
obtainable through the combination of two forms of transverse
momentum distributions at low and high transverse momentum
regions, Eqs. (\ref{d2NsdpT2low}) and (\ref{d2NsdpT2high}).

Here, the same transverse momentum distribution for both strange
and anti-strange quarks is assumed as the baryon chemical
potential at the LHC $\sqrt{s_{NN}}=5.02$ TeV energy is almost
zero. For the $\phi$ meson size, $\sigma_{\phi}$, the 0.46 fm is
taken as the root-mean-square radius of the $\phi$ meson
\cite{Chen:2006vc, Wang:2023wlq}, corresponding to the oscillator
frequency of the $\phi$ meson, $\omega_{\phi}=1.104$ GeV. For the
degeneracy factor of the $\phi$ meson, $g_{\phi}=3/(2\cdot3)^2$ is
taken.

Regarding the coalescence volume at the LHC $\sqrt{s_{NN}}=5.02$
TeV energy, we evaluate it based on the entropy conservation
condition between the critical $T_C$ and hadronization $T_H$
temperatures, $s_C V_C=s_H V_H$ where $s_C$ and $s_H$ are the
entropy densities, $V_C$ and $V_H$ volumes at critical and
hadronization temperatures in the system, respectively. Starting
from the condition at hadronization in the statistical
hadronization model, $T_H=156.5$ MeV, $V_H=4997$ fm$^3$
\cite{Andronic:2021erx}, we trace back in temperatures to find the
coalescence volume, or the volume at critical temperature
$T_C=166$ MeV \cite{Cho:2017dcy}. The data on entropy densities
$s=(\epsilon+p)/T$ at different temperatures are available through
the combination between the pressure $p(T)/T^4$ and trace anomaly,
$(\epsilon-3p)(T)/T^4$ from the Lattice calculation
\cite{Borsanyi:2010cj}. The coalescence volume at
$\sqrt{s_{NN}}=5.02$ TeV is evaluated to be 3360 fm$^3$, a little
bit smaller than that at the LHC $\sqrt{s_{NN}}=2.76$ TeV energy,
3530 fm$^3$ \cite{Cho:2017dcy} because of the smaller
hadronization volume 4997 fm$^3$ at $\sqrt{s_{NN}}=5.02$ TeV
compared to 5380 fm$^3$ at $\sqrt{s_{NN}}=2.76$ TeV
\cite{Stachel:2013zma}.

Using the coalescence volume 3360 fm$^3$ with the mass of strange
quarks in the quark-gluon plasma 500 MeV, we can now obtain the
information on the transverse momentum spectrum of strange quarks
in the quark-gluon plasma. First, by plugging the combined
transverse momentum distributions of strange quarks in Eqs.
(\ref{d2NsdpT2low}) and (\ref{d2NsdpT2high}) with arbitrary
parameters into those in Eq. (\ref{CoalTransphi}), we calculate
the transverse momentum distribution of the $\phi$ meson. Then, we
compare the result to the experimental measurement at the LHC
\cite{ALICE:2019xyr}, adjust the parameters in transverse momentum
distribution of strange quarks, and repeat the calculation until
we finally obtain the information on the transverse momentum
spectrum of strange quarks which can explain the transverse
momentum distribution of the $\phi$ meson measured by the ALICE
Collaboration. With this approach, we can determine parameters in
Eqs. (\ref{d2NsdpT2low}) and (\ref{d2NsdpT2high}).

\begin{table}[!t]
\caption{The parameters in the transverse momentum distribution of
strange quarks at $\sqrt{s_{NN}}=5.02$ TeV, Eq.
(\ref{d2NsdpT2high}). } \label{parameters_strange}
\begin{center}
\begin{tabular}{c|c|c|c}
\hline \hline
$s_1$ (GeV$^{-2}$) & $s_2$ & $s_3$ & $s_4$ (GeV$^{-2}$) \\
\hline 21.95 & 0.17 & 3.23 & 80112 \\
\hline
$s_5$ & $s_6$ & $p_{0T}$ (GeV) & $p_{sT}'$ (GeV)\\
\hline 0.65 & 10.29 & 1.0 & 1.5 \\
\hline \hline
\end{tabular}
\end{center}
\end{table}

\begin{figure}[!t]
\begin{center}
\includegraphics[width=0.50\textwidth]{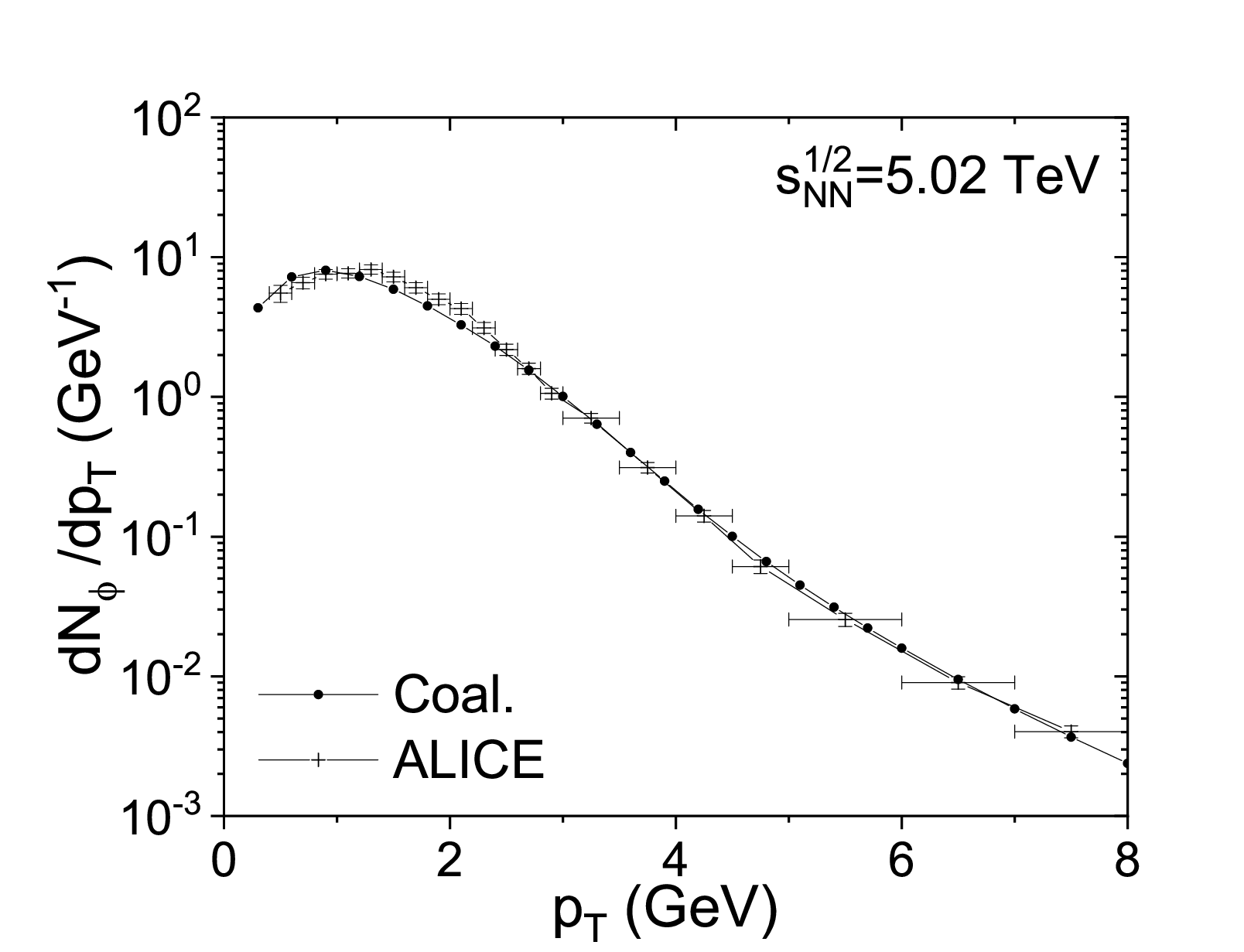}
\end{center}
\caption{Transverse momentum distribution of the $\phi$ meson,
$dN_{\phi}/dp_T$ at mid-rapidity at $\sqrt{s_{NN}}=5.02$ TeV
together with the measurement of the $\phi$ meson at the same LHC
energy by ALICE Collaboration \cite{ALICE:2019xyr}. }
\label{pTdistribution_phi}
\end{figure}

We show in Fig. \ref{pTdistribution_phi} the transverse momentum
distribution of the $\phi$ meson together with the measurement by
the ALICE Collaboration \cite{ALICE:2019xyr}. The effective
temperature of the strange quark in Eq. (\ref{d2NsdpT2low}) is
found to be, $T_{eff}$=173 MeV, and the parameters in Eq.
(\ref{d2NsdpT2high}) are summarized in Table
\ref{parameters_strange}. We find that the total number of strange
quarks at mid-rapidity at $\sqrt{s_{NN}}=5.02$ TeV,
$dN_s/dy|_{|y|<0.5}$ is about 784 after integrating the
combination of two forms of strange quark transverse momentum
distributions at low and high transverse momentum regions, Eqs.
(\ref{d2NsdpT2low}) and (\ref{d2NsdpT2high}) over the entire
transverse momentum regions.

Here, major feed-down contributions from charm-strange mesons to
the production of $\phi$ mesons \cite{ParticleDataGroup:2024cfk}
are taken into account in the evaluation of the transverse
momentum distribution of the $\phi$ meson,
\begin{eqnarray}
&& \frac{dN_{\phi}^{all}}{dp_T}=
\frac{dN_{\phi}}{dp_T}+0.157~\frac{dN_{D_s}^{all}}{dp_T},
\nonumber \\
&& \frac{dN_{D_s}^{all}}{dp_T}= \frac{dN_{D_s}}{dp_T}
+\frac{dN_{D_s^*}}{dp_T}+\frac{dN_{D_{s0}^*(2317)}}{dp_T}
\nonumber \\
&& \qquad\quad~+0.820~\frac{dN_{D_{s1}(2460)}}{dp_T}
+0.0802\frac{dN_c^{Frag}}{dp_T}, \label{feeddown}
\end{eqnarray}
but those contributions to the $\phi$ meson are found to be
negligible compared to $\phi$ mesons purely produced from strange
quarks in the quark-gluon plasma by coalescence.

\subsection{Charm quark transverse momentum distributions}

In order to obtain the transverse momentum distribution of charm
quarks, we consider that of $D^0$ mesons formed from charm and
light quarks in the quark-gluon plasma by coalescence. Thus, it is
also necessary to get information on the transverse momentum
distribution of light quarks. To this end, on one hand, we adopt a
similar transverse momentum distribution for light quarks as that
of strange quarks at low $p_T$ regions after replacing the
effective temperature and constituent quark mass. As the light
quark is affected more by the flow than the strange quark, the
effective temperature in the light quark transverse momentum
distribution at low $p_T$ regions is expected to be larger than
that of the strange quark. On the other hand, we take the exactly
same transverse momentum distribution as that of strange quarks
for light quarks at high $p_T$ regions. Due to the smaller masses
of light and strange quarks compared to the transverse momenta at
high $p_T$ regions, it is expected that the transverse momentum
distributions of light and strange quarks show the same behavior
at high $p_T$ regions.

For the transverse momentum distribution of charm quarks
$\sqrt{s_{NN}}=5.02$ TeV, we take the same form of the transverse
momentum distribution of charm quarks as that already developed at
$\sqrt{s_{NN}}=2.76$ TeV \cite{Plumari:2017ntm},

\begin{eqnarray}
&& \frac{d^2N_c}{d^2\vec p_{cT}}=\left\{
\begin{array}{ll}
c_1e^{(-c_2 (\frac{p_{cT}}{p_{0T}})^{c_3})} & \quad p_{cT} \le p_{cT}' \\
c_4e^{(-c_5 (\frac{p_{cT}}{p_{0T}})^{c_6})} \nonumber \\
+\frac{c_7}{\Big(1.0+(\frac{p_{cT}}{p_{0T}})^{c_8}\Big)^{c_9}} &
\quad p_{cT} > p_{cT}'
\end{array} \right. \\
\label{dNcdpT}
\end{eqnarray}
with parameters $c_1, c_2, c_3, c_4, c_5, c_6, c_7, c_8$ and
$c_9$, which will be determined later.

In addition to the transverse momentum distributions of quarks, it
is also required to determine the oscillator frequency for the
harmonic oscillator wave functions, $\omega=1/\mu\sigma^2$, when
computing the transverse momentum distribution of charmed hadrons
produced by coalescence. Here, we determine the oscillator
frequency on the condition that the number of charm quarks
produced by initial hard scattering in relativistic heavy ion
collisions should be conserved in the entire process, even after
chemical freeze-out. Reminding that charmed hadrons are dominantly
produced by recombination mostly at low $p_T$ regions, we require
that all charm quarks at zero $p_T$ should be converted entirely
into charmed hadrons, leading to the number of charm quarks at
zero $p_T$ equal to the sum of yields of all charmed hadrons
formed by coalescence at zero $p_T$ \cite{Oh:2009zj, Cho:2019syk}.

Here, we consider only the singly charmed mesons and baryons,
including six charm mesons, $D$, $D^*$, $D_s$, $D_s^*$,
$D_{s0}^*(2317)$, and $D_{s1}(2460)$ and ten charm baryons,
$\Lambda_c$, $\Sigma_c(2455)$, $\Sigma_c(2520)$,
$\Lambda_c(2595)$, $\Lambda_c(2625)$, $\Xi_c$, $\Xi_c'$,
$\Xi_c(2645)$, $\Omega(2695)$, and $\Omega(2770)$ in determining
the oscillator frequency for the formation of charmed hadrons. We
find that contributions to the total yield of charmed hadrons at
zero $p_T$ from hidden charmed mesons as well as multicharmed
hadrons containing more than two charm quarks are negligible
compared to those from the singly charmed hadrons mentioned above.

As the production of hadrons by coalescence is dominant at low
$p_T$ regions, it is necessary to take the other production
mechanism, in particular at high $p_T$ regions, i.e., hadron
production by fragmentation into account in order to explain more
exactly the transverse momentum distribution of hadrons, here the
$D^0$ meson in the entire transverse momentum region. We,
therefore, first evaluate as functions of charm quark transverse
momenta the amount of charm quarks consumed by all charmed hadrons
mentioned above when they are produced from charm and light quarks
by coalescence, and then let the remaining charm quarks be fully
converted into designated charmed hadrons according to the
fragmentation fraction \cite{Lisovyi:2015uqa}.

We adopt here the Peterson fragmentation function
\cite{Peterson:1982ak} with $\epsilon_c=0.022$ to calculate the
transverse momentum distribution of $D^0$ mesons produced through
fragmentation. The feed-down contribution to the production of
$D^0$ meson by fragmentation is also taken into account here,
implying that 60.86$\%$ of the remaining charm quarks contribute
to $D^0$ meson production directly or indirectly via $D^*$ by
fragmentation \cite{Lisovyi:2015uqa}.

\begin{figure}[!t]
\begin{center}
\includegraphics[width=0.50\textwidth]{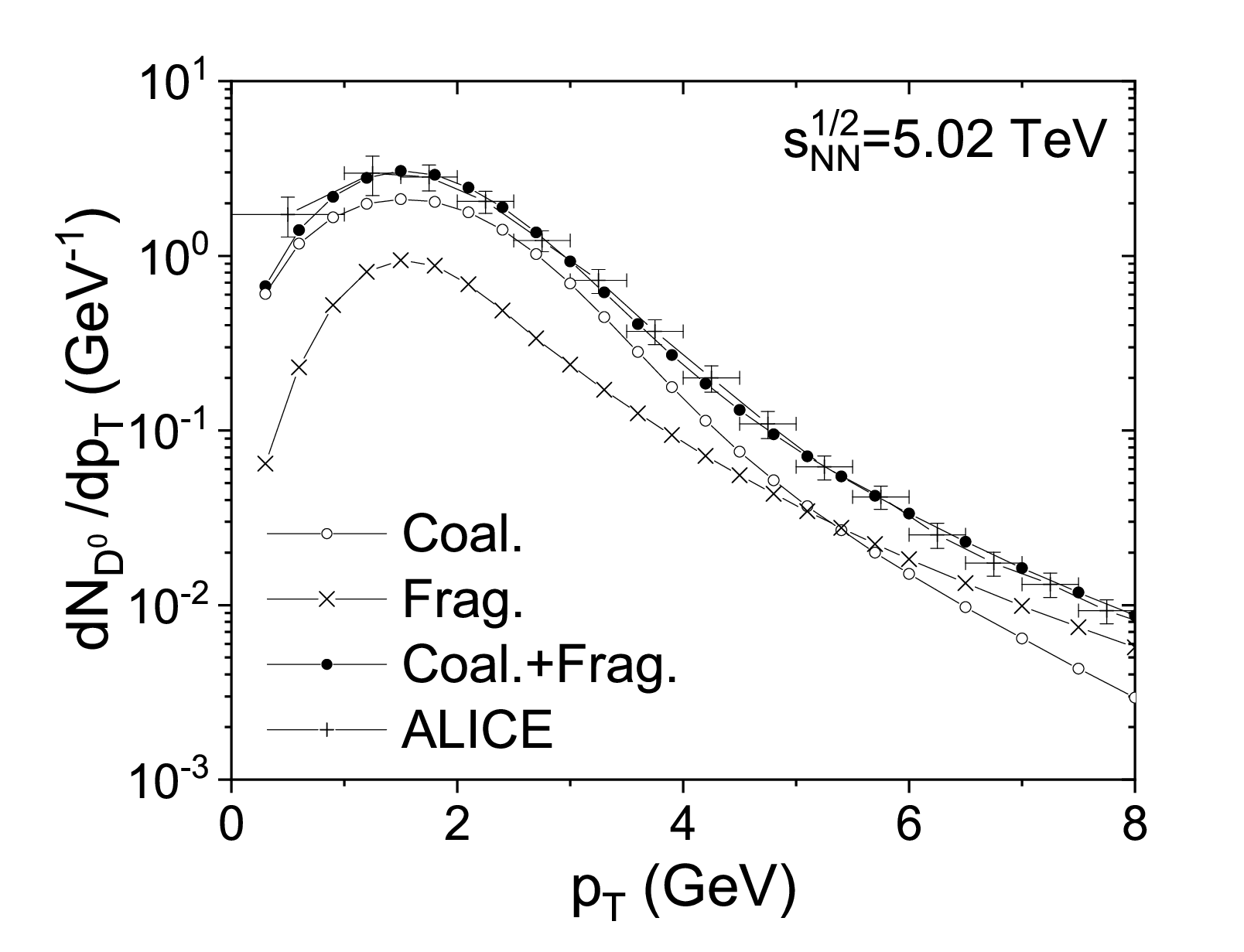}
\end{center}
\caption{Transverse momentum distribution of $D^0$ mesons at
mid-rapidity, $dN_{D^0}/dp_T$ at the LHC $\sqrt{s_{NN}}=5.02$ TeV
energy together with the experimental measurement of that by ALICE
Collaboration \cite{ALICE:2021rxa} at 0-10$\%$ centrality in Pb-Pb
collisions at $\sqrt{s_{NN}}=5.02$ TeV. Also shown are transverse
momentum distributions of $D^0$ mesons produced purely by
coalescence and also by fragmentation. Here, feed-down
contributions to the production of $D^0$ mesons, Eq.
(\ref{feeddownD0}) have been taken into account. }
\label{pTdistribution_D0}
\end{figure}

Using the same method employed to obtain the transverse momentum
distribution of strange quarks, we derive the transverse momentum
distribution of charm quarks by repeatedly comparing our results
with the experimental measurement of the $D^0$ meson transverse
momentum distribution by ALICE Collaboration at
$\sqrt{s_{NN}}=5.02$ at the LHC. We show in Fig.
\ref{pTdistribution_D0} the transverse momentum distribution of
$D^0$ mesons produced by coalescence and also by fragmentation,
together with the total transverse momentum distribution of $D^0$
mesons. Also shown in Fig. \ref{pTdistribution_D0} is the
experimental measurement of $D^0$ transverse momentum distribution
by ALICE Collaboration at 0-10$\%$ centrality in Pb-Pb collisions
at $\sqrt{s_{NN}}=5.02$ \cite{ALICE:2021rxa}. Here we have taken
into account feed-down contributions for the $D^0$
\cite{ParticleDataGroup:2024cfk, Lisovyi:2015uqa}, i.e., we
consider the following for the feed-down contribution to the
production of $D^0$ mesons,

\begin{eqnarray}
&& \frac{dN_{D^0}^{all}}{dp_T}=
\frac{dN_{D^0}}{dp_T}+\frac{dN_{D^{*0}}}{dp_T},
\nonumber \\
&& \qquad\quad~+0.677\frac{dN_{D^{*+}}}{dp_T}+0.6086
\frac{dN_c^{Frag}}{dp_T}. \label{feeddownD0}
\end{eqnarray}
As shown in Fig. \ref{pTdistribution_D0}, both production
mechanisms, production of the $D^0$ by coalescence and by
fragmentation, play essential roles for the $D^0$ meson production
in the entire transverse momentum region. While the production of
$D^0$ mesons by coalescence is dominant at low $p_T$ regions, up
to the regions of about $p_T<5$ GeV, the $D^0$ meson is more
produced by fragmentation than by coalescence at high $p_T$
regions, above the regions of about $p_T>5$ GeV.

We summarize in Table \ref{parameters_charm} the parameters of the
charm quark transverse momentum distribution in Eq.
(\ref{dNcdpT}). We take for some of the parameters, $c_4$, $c_5$,
and $c_6$, the same values already obtained in the transverse
momentum distribution of charm quarks at $\sqrt{s_{NN}}=2.76$ TeV
\cite{Plumari:2017ntm}. In determining the other parameters shown
in Table \ref{parameters_charm}, the constituent quark mass 300
MeV with the effective temperature $T_{eff}=187$ MeV for light
quarks is taken for the transverse momentum distribution of light
quarks at low $p_T$ regions. In addition, we find the oscillator
frequency $\omega=0.103$ GeV, which enables all charm quarks at
zero $p_T$ to be fully converted to charmed hadrons when those
charmed hadrons are produced by quark coalescence at zero $p_T$ in
the quark-gluon plasma. Adopting this oscillator frequency results
in $\sqrt{\langle r^2\rangle}=1.51$ fm for the root-mean-square
radius of the $D^0$ meson.

\begin{table}[!t]
\caption{The parameters in the transverse momentum distribution of
charm quarks at $\sqrt{s_{NN}}=5.02$ TeV, Eq. (\ref{dNcdpT}). }
\label{parameters_charm}
\begin{center}
\begin{tabular}{c|c|c|c|c}
\hline \hline
$c_1$ (GeV$^{-2}$) & $c_2$ & $c_3$ & $c_4$ (GeV$^{-2}$) & $c_5$ \\
\hline 1.63 & 0.27 & 2.03 & 7.95 & 3.49 \\
\hline
$c_6$ & $c_7$ (GeV$^{-2}$) & $c_8$ & $c_9$ & $p_{cT}'$ (GeV)\\
\hline 3.59 & 90112 & 0.50 & 14.19 & 1.80 \\
\hline \hline
\end{tabular}
\end{center}
\end{table}
With the above transverse momentum distribution of charm and light
quarks, we can obtain the total number of charm and light quarks
available in the quark-gluon plasma at $\sqrt{s_{NN}}=5.02$ TeV.
By performing the integration of the transverse momentum
distribution of charm and light quarks over all transverse
momenta, we find $dN_c/dy=17.4$ and $dN_q/dy=1040$ at mid-rapidity
for the total number of charm and light quarks, respectively,
which is approximately 1.6 and 1.5 times larger than those
expected at $\sqrt{s_{NN}}=2.76$ TeV \cite{Cho:2017dcy}.

\section{Yields and transverse momentum distributions of
charm-strange mesons}

\subsection{Transverse momentum distributions of charm-strange
mesons}

Now we calculate the transverse momentum distributions of $D_s$,
$D_s^*$, $D_{s0}^*(2317)$, and $D_{s1}(2460)$ mesons, Eqs.
(\ref{CoalTransCSs}) and (\ref{CoalTransCSp}) when they are
produced from charm and strange quarks by quark recombination
using the transverse momentum distributions of strange and charm
quarks, Eqs. (\ref{d2NsdpT2low}), (\ref{d2NsdpT2high}), and
(\ref{dNcdpT}) with the oscillator frequency 0.103 GeV and the
coalescence volume 3360 fm$^3$ at $\sqrt{s_{NN}}=5.02$ TeV.

\begin{figure}[!t]
\begin{center}
\includegraphics[width=0.50\textwidth]{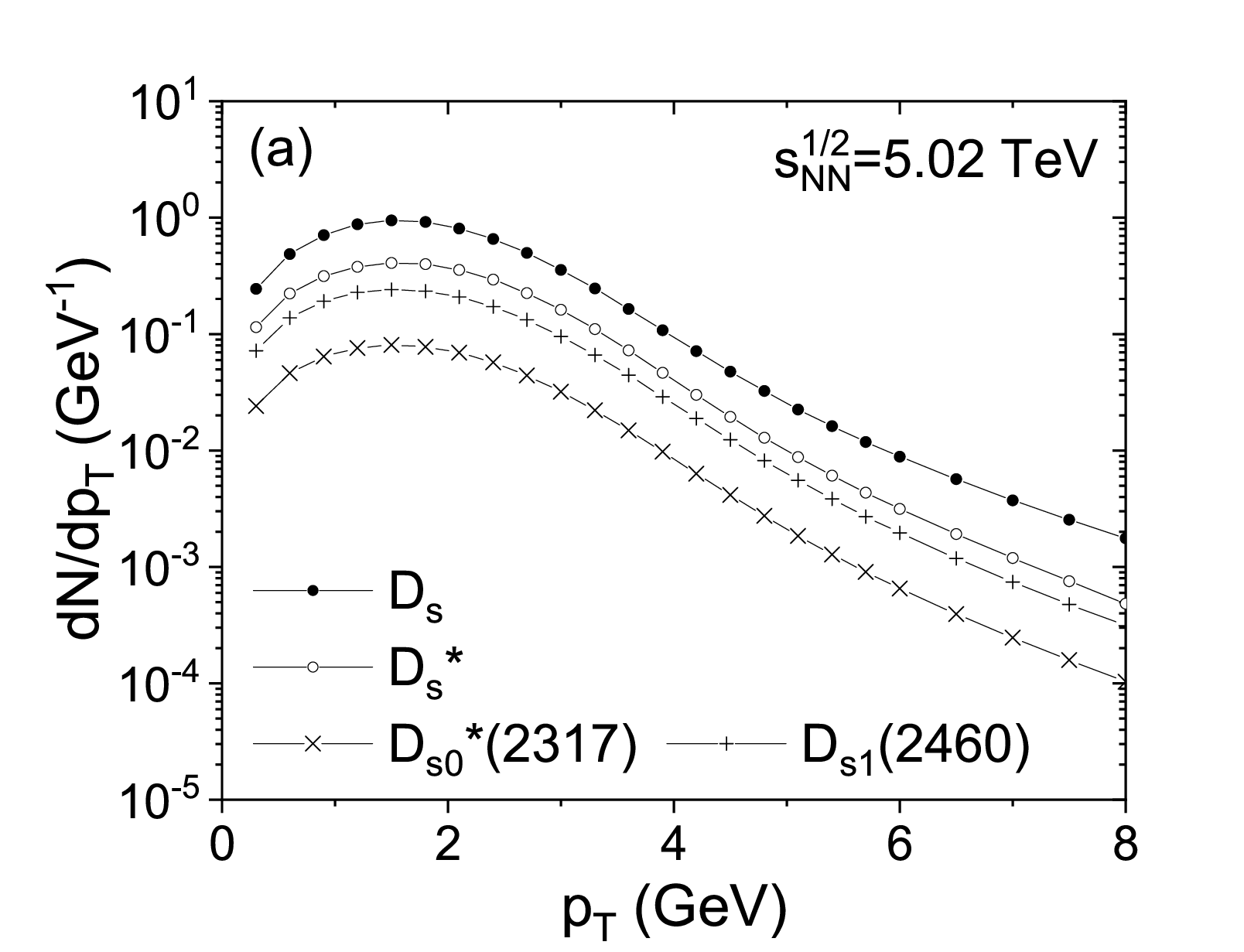}
\includegraphics[width=0.50\textwidth]{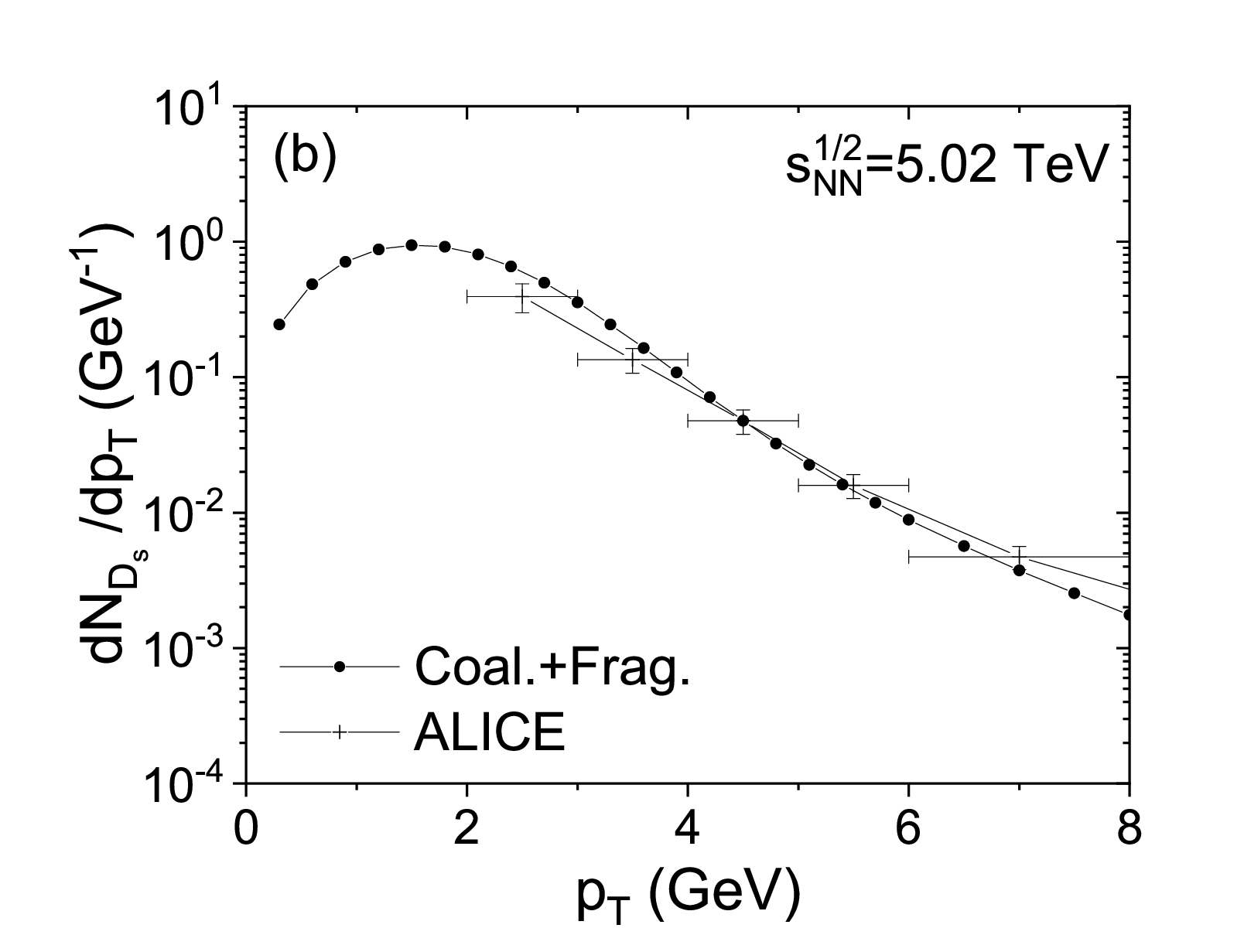}
\end{center}
\caption{(a) Transverse momentum distributions of $D_s$, $D_s^*$,
$D_{s0}^*(2317)$, and $D_{s1}(2460)$ mesons at mid-rapidity,
$dN/dp_T$ at $\sqrt{s_{NN}}=5.02$ TeV and (b) transverse momentum
distribution of $D_s$ mesons, $dN_{D_s}/dp_T$ at mid-rapidity at
$\sqrt{s_{NN}}=5.02$ TeV with that of the $D_s$ meson measured at
the same rapidity range and energy by ALICE Collaboration
\cite{ALICE:2021kfc}. For the transverse momentum distribution of
the $D_s$ meson, the feed-down contribution from $D_s^*$,
$D_{s0}^*(2317)$, and $D_{s1}(2460)$ mesons to the $D_s$ as well
as the production of $D_s$ meson by fragmentation from the
remaining charm quarks, Eq. (\ref{feeddown}) have been taken into
account.} \label{pTdistribution_charmstrange}
\end{figure}
We show in Fig. \ref{pTdistribution_charmstrange}(a) the
transverse momentum distributions, $dN/dp_T$ of charm-strange
mesons, the $D_s$, $D_s^*$, $D_{s0}^*(2317)$, and $D_{s1}(2460)$
at mid-rapidity at LHC $\sqrt{s_{NN}}=5.02$ TeV energy. Also shown
in Fig. \ref{pTdistribution_charmstrange}(b) are the transverse
momentum distribution of the $D_s$ meson $dN_{D_s}/dp_T$ at
mid-rapidity together with that of the $D_s$ meson measured at the
same rapidity range and energy by ALICE Collaboration
\cite{ALICE:2021kfc}. For the transverse momentum distribution of
the $D_s$ meson in Fig. \ref{pTdistribution_charmstrange}, the
feed-down contribution from $D_s^*$, $D_{s0}^*(2317)$, and
$D_{s1}(2460)$ mesons to the $D_s$ as well as the production of
$D_s$ meson by fragmentation from the remaining charm quarks, Eq.
(\ref{feeddown}) have been taken into account. Here, we assume the
same transverse momentum distribution of the daughter meson, the
$D_s$ as those of $D_s^*$, $D_{s0}^*(2317)$, and $D_{s1}(2460)$
mesons after the decay from the $D_s^*$, $D_{s0}^*(2317)$, and
$D_{s1}(2460)$ to the $D_s$ meson as the $D_s$ meson is much
heavier than the other daughter meson, mostly the pion in the
decay process.

As shown in Fig. \ref{pTdistribution_charmstrange}(b), the
transverse momentum distribution of the $D_s$ meson agree
reasonable well with the experimental measurement of the $D_s$
meson transverse momentum distribution by ALICE Collaboration
\cite{ALICE:2021kfc}.

\subsection{Yields of charm-strange mesons}

We calculate the yield of charm-strange mesons mentioned above, as
well as the $D^0$ and $\phi$ mesons, by performing the integration
of transverse momentum distributions shown in Figs.
\ref{pTdistribution_phi}, \ref{pTdistribution_D0}, and
\ref{pTdistribution_charmstrange} over all transverse momenta.
Here we have taken into account feed-down contributions for the
$\phi$, $D^0$ and $D_s$, denoted by and $\phi^{all}$, $D^{0 all}$
and $D_s^{all}$, respectively, using Eqs. (\ref{feeddown}) and
(\ref{feeddownD0}) \cite{ParticleDataGroup:2024cfk,
Lisovyi:2015uqa}.

Moreover, we calculate the yield of the above mesons assuming that
they are in thermal equilibrium at chemical freeze-out. The yield
in the thermal model for the charmed hadron is given by,

\begin{equation}
N^\mathrm{Ther.} = g V_c \int \frac{d \vec p}{(2\pi)^3}
\frac{1}{\gamma_c^{-n}e^{\sqrt{p^2+m^2}/T_c}\pm 1}, \label{NTher}
\end{equation}
with $g$ being the degeneracy factor of a hadron, $V_c$ and $T_c$
the volume and temperature at chemical freeze-out, respectively.
In Eq. (\ref{NTher}), $m$ is the mass of the hadron, and
$\gamma_c$ is the fugacity of charm quarks with the number of
charm quarks $n$ in charmed hadrons. The fugacity of charm quark,
$\gamma_c$, is introduced here to reflect charm quarks in
non-equilibrium because of their heavier masses compared to
energies available in a quark-gluon plasma in heavy ion
collisions.

We adopt here the volume, $V_c=4997$ fm$^3$, and temperature
$T_c=156.5$ MeV at chemical freeze-out from the statistical
hadronization model analysis at $\sqrt{s_{NN}}=5.02$ TeV at
mid-rapidity in most central collisions, 0-10$\%$ centrality
regions \cite{Andronic:2021erx}. For the charm fugacity
$\gamma_c$, we use the number of charm quarks at mid-rapidity,
$dN_c/dy|_{|y|<0.5}=17.4$ obtained in the previous section for the
total numbers of charm quarks produced from initial hard
collisions in heavy ion collisions, and compute the charm fugacity
by requiring that all charm quarks produced initially should be
conserved via the distribution of those charm quarks to all
charmed hadrons after chemical freeze-out \cite{Cho:2010db,
Cho:2011ew, Cho:2017dcy, Cho:2019syk}. We consider here only
singly charmed mesons and baryons introduced in the previous
section when evaluating the oscillator frequency of the Wigner
function $\omega$ under the condition that the number of charm
quarks at zero $p_T$ should be equal to the sum of yields of all
charmed hadrons formed by coalescence at zero $p_T$; six charmed
mesons, $D$, $D^*$, $D_s$, $D_s^*$, $D_{s0}^*(2317)$, and
$D_{s1}(2460)$ and ten charm baryons, $\Lambda_c$,
$\Sigma_c(2455)$, $\Sigma_c(2520)$, $\Lambda_c(2595)$,
$\Lambda_c(2625)$, $\Xi_c$, $\Xi_c'$, $\Xi_c(2645)$,
$\Omega(2695)$, and $\Omega(2770)$.

By equating the sum of the yields of the charmed hadrons mentioned
above at chemical freeze-out to be the number of charm quarks in
the system, 17.4, we calculate $\gamma_c=47.8$ for the fugacity of
charm quarks. Then, we evaluate the yield of charm-strange mesons
using Eq. (\ref{NTher}). The feed-down contributions to the
production of $\phi$, $D^0$, and $D_s$ mesons have been taken into
account, respectively, based on Eqs. (\ref{feeddown}) and
(\ref{feeddownD0}) without considering the production of those
mesons by fragmentation \cite{ParticleDataGroup:2024cfk}. We
summarize our results in Table. \ref{charmestrangeyields}.

We also show in Table \ref{charmestrangeyields} yields of the
$\phi$, $D^0$ and $D_s$ meson experimentally measured at
mid-rapidity, $|y|<0.5$ in 0-10$\%$ centrality class at LHC in
$\sqrt{s_{NN}}=5.02$ TeV Pb+Pb collisions by ALICE Collaboration
\cite{ALICE:2021kfc, ALICE:2021rxa, ALICE:2021ptz}. Listed in
Table \ref{charmestrangeyields} are also yields of $D^0$ and $D_s$
mesons evaluated in the Statistical Hadonization Model with charm
(SHMc) \cite{Andronic:2021erx}.

\begin{table}[!t]
\caption{Yields of the $\phi$ and $D^0$ meson as well as those of
charm-strange mesons, the $D_s$, $D_s^*$, $D_{s0}^*(2317)$, and
$D_{s1}(2460)$ at mid-rapidity obtained by integrating the
transverse momentum distributions shown in Figs.
\ref{pTdistribution_phi}, \ref{pTdistribution_D0}, and
\ref{pTdistribution_charmstrange} over all transverse momenta.
Also listed are yields of mesons mentioned above evaluated at
chemical freeze-out assuming that they are in thermal equilibrium,
Eq. (\ref{NTher}), and yields of the $\phi$, $D^0$ and $D_s$ meson
experimentally measured at mid-rapidity, $|y|<0.5$ in 0-10$\%$
centrality class at LHC in $\sqrt{s_{NN}}=5.02$ TeV Pb+Pb
collisions by ALICE Collaboration \cite{ALICE:2021kfc,
ALICE:2021rxa, ALICE:2021ptz}. Yields of $D^0$ and $D_s$ mesons in
the Statistical Hadronization Model with charm (SHMc)
\cite{Andronic:2021erx} are also presented for comparison.}
\label{charmestrangeyields}
\begin{center}
\begin{tabular}{c|c|c|c|c}
\hline \hline
Yields & Coal. & Ther. & ALICE  & SHMc  \\
\hline $\phi^{all}$ & ~14.3~ & ~15.8~ & 14.937 \cite{ALICE:2021ptz} &  \\
$D^{0 all}$ & ~6.54~ & ~7.84~  &  6.819 \cite{ALICE:2021rxa} & 6.42 \cite{Andronic:2021erx}  \\
$D_s$ & ~0.321~ & ~1.34~ & &  \\
$D_s^*$ & ~0.963~ & ~1.77~ & &  \\
$D_{s0}^*(2317)$ & ~0.192~ & ~0.180~ & &   \\
$D_{s1}(2460)$ & ~0.576~ & ~0.237~ & & \\
$D_s^{all}$ & ~2.18~ & ~3.49~  & 1.89 \cite{ALICE:2021kfc} & 2.22 \cite{Andronic:2021erx} \\
\hline \hline
\end{tabular}
\end{center}
\end{table}

The yields of charm-strange mesons evaluated from the integration
of the transverse momentum distributions shown in Figs.
\ref{pTdistribution_phi}, \ref{pTdistribution_D0}, and
\ref{pTdistribution_charmstrange} over all transverse momentum
regions show reasonable agreements with the experimental
measurements. The yields of those mesons based on the assumption
of thermal equilibrium are found to be larger than measurements,
as well as those obtained in the coalescence model. This must be
attributable to a rough assumption on the complete thermal
equilibrium of charmed hadrons without considering corona effects
at high transverse momentum regions. In addition, the larger
fugacity owing to the limited number of charmed hadrons, a total
of 16 charmed hadrons included in its evaluation results in the
relatively larger yields compared to the measurements.

\subsection{Transverse momentum distribution ratio between
$D_s$ and $D^0$ mesons}

Based on transverse momentum distributions of $D^0$ and $D_s$
mesons as shown in Fig. \ref{pTdistribution_D0} and Fig.
\ref{pTdistribution_charmstrange}(b) we evaluate the transverse
momentum distribution ratio between those two mesons. The
transverse momentum distribution ratio between the $D_s$ and $D^0$
is the ratio between a charm-light and a charm-strange meson,
$c\bar{q}/c\bar{s}$. After cancelling out the effects from the
same charm flavor appearing in the numerator and denominator, one
can obtain information on the roles of light and strange quarks in
the process of hadronization.

\begin{figure}[!t]
\begin{center}
\includegraphics[width=0.50\textwidth]{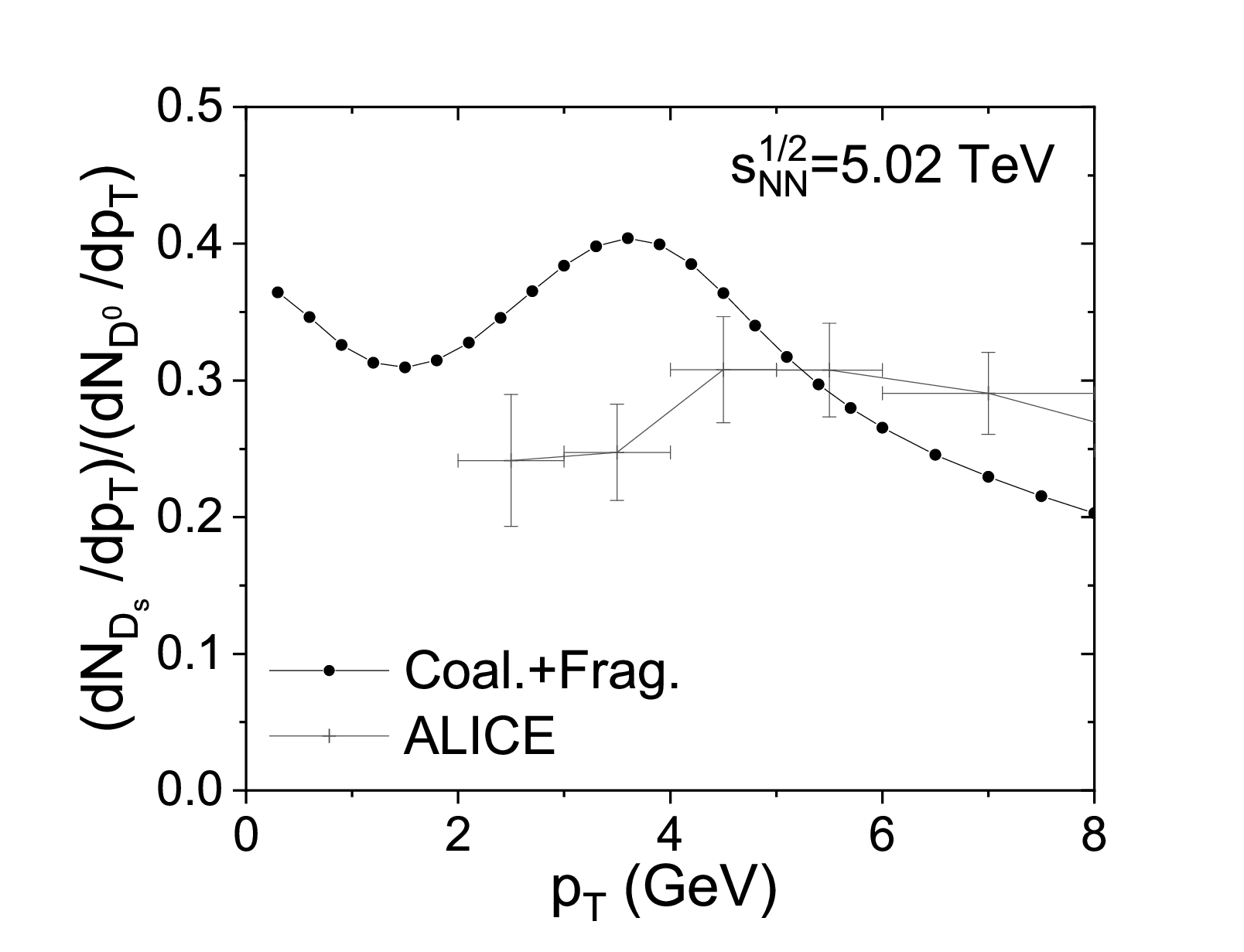}
\end{center}
\caption{Transverse momentum distribution ratio between $D_s$ and
$D^0$ mesons, $(dN_{D_s}/dp_T)/(dN_{D^0}/dp_T)$ at mid-rapidity in
$0-10\%$ centrality at $\sqrt{s_{NN}}=5.02$ TeV at LHC. Also shown
is the same transverse momentum distribution ratio between $D_s$
and $D^0$ mesons measured by ALICE Collaboration at
$\sqrt{s_{NN}}=5.02$ TeV at LHC \cite{ALICE:2021kfc}.}
\label{pTdistributionratioDstoD0}
\end{figure}

We show in Fig. \ref{pTdistributionratioDstoD0} the transverse
momentum distribution ratio between the $D_s$ and $D^0$ meson,
$(dN_{D_s}/dp_T)/(dN_{D^0}/dp_T)$ at mid-rapidity in $0-10\%$
centrality at $\sqrt{s_{NN}}=5.02$ TeV at LHC. Also shown in Fig.
\ref{pTdistributionratioDstoD0} is the same transverse momentum
distribution ratio between $D_s$ and $D^0$ mesons measured by
ALICE Collaboration at the same LHC energy \cite{ALICE:2021kfc}.
Here, the feed-down as well as the fragmentation contributions to
both mesons have been taken into account, Eqs. (\ref{feeddownD0})
and (\ref{feeddown}).

As shown in Fig. \ref{pTdistributionratioDstoD0}, the transverse
momentum distribution ratio between $D_s$ and $D^0$ mesons
evaluated in this work is larger at low transverse momentum
regions compared to the measurement by ALICE Collaboration
\cite{ALICE:2021kfc}, while it decreases from about $p_T=4$ GeV,
and becomes lower than the measurement of that at $p_T=5$ GeV. The
ratio lies between 0.2 and 0.4, around the yield ratio between
$D_s$ and $D^0$ mesons, $2.18/6.54\approx 0.33$, Table
\ref{charmestrangeyields}.

The ratio exhibits a peak at about 4 GeV, similar to the result in
the Statistical Hadronization Model with charm (SHMc)
\cite{Andronic:2021erx}. The peak in the ratio is mainly
attributable to the mass difference between $D_s$ and $D^0$
mesons, specifically the mass difference between $s$ and $u$
quarks; the coalescence probability for forming $D^0$ and $D_s$
mesons has different transverse momentum dependence due to
different masses of $u$ and $s$ quarks.

The momentum of heavy quark hadrons is mainly governed by that of
the heavy quark. Accordingly, the momentum of a light quark in the
$D^0$ meson occupies roughly a sixth of the total momentum of the
$D^0$ meson, as the constituent mass of charm quarks is five times
heavier than that of light quarks. On the other hand, the momentum
of a strange quark in the $D_s$ meson is about a quarter of the
total momentum of the $D_s$ meson because the constituent mass of
charm quarks is three times heavier than that of strange quarks.
Therefore, the light and strange quarks with the same transverse
momenta contribute to the production of $D^0$ and $D_s$ mesons
differently, resulting in the transverse momentum of $D^0$ mesons
higher than that of $D_s$ mesons by 1.5 times. These differences
cause the coalescence probabilities to depend differently on the
transverse momenta, resulting in a peak in the transverse momentum
distribution ratio between $D_s$ and $D^0$ mesons. In other words,
the light and strange quarks moving collectively with the same
transverse momenta in thermal equilibrium lead to different
transverse momentum distributions for $D^0$ and $D_s$ mesons.

\subsection{$D_{s0}^*(2317)$ meson in a four quark state}

Here, we evaluate the yield and transverse momentum distribution
of the $D_{s0}^*(2317)$ meson by assuming that it exists in a
four-quark state. Thus, the $D_{s0}^*(2317)$ meson is considered
to be formed from two light, one strange, and one charm quark in
quark-gluon plasma by quark coalescence. The production of the
$D_{s0}^*(2317)$ meson when it is in both two- and four-quark
states was studied about two decades ago for the case when it is
produced at RHIC at $\sqrt{s_{NN}}=200$ GeV Au+Au collisions
\cite{Chen:2007zp}, focusing more on the interaction of the
$D_{s0}^*(2317)$ meson with light hadrons during the hadronic
stage in heavy ion collisions. Thus, it is necessary to update the
investigation on the production of the $D_{s0}^*(2317)$ meson,
providing its yield and transverse momentum distribution at
$\sqrt{s_{NN}}=5.02$ TeV in Pb+Pb collisions also for the
$D_{s0}^*(2317)$ meson in a four-quark state. Compared to relative
$p$-wave state between charm and strange quarks for the
$D_{s0}^*(2317)$ meson in a two-quark state, all quarks inside the
$D_{s0}^*(2317)$ in a four-quark state exist in a $s$-wave to
satisfy its spin and parity, $J^p=0^+$.

The transverse momentum distribution of $D_{s0}^*(2317)$ meson in
a four-quark state can be obtained from that of the $X(3872)$
meson in a four-quark state \cite{Cho:2019syk} replacing one charm
quark with one strange quark,

{\allowdisplaybreaks
\begin{eqnarray}
 && \frac{d^2N_{D_{s0}^*(2317)}}{d^2\vec p_T}=\frac{g_{D_{s0}^*(2317)}}
 {V^3}(2\sqrt{\pi})^9 (\sigma_1\sigma_2\sigma_3)^3 \nonumber \\
&& \qquad\qquad\quad \times\int d^2\vec p_{qT}d^2\vec
p_{\bar{q}T}d^2\vec p_{cT}d^2\vec p_{\bar{s}T} \nonumber \\
&& \qquad\qquad\quad \times \frac{d^2N_q}{d^2 \vec p_{qT}}
\frac{d^2 N_{\bar{q}}}{d^2\vec p_{\bar{q}T}} \frac{d^2N_c}{d^2\vec
p_{cT}} \frac{d^2N_{\bar{s}}}{d^2\vec p_{\bar{s}T}} \nonumber \\
&& \qquad\qquad\quad \times\delta^{(2)}(\vec p_T-\vec p_{qT}-\vec
p_{\bar{q}T}-\vec p_{cT}-\vec p_{\bar{s}T}) \nonumber \\
&& \qquad\qquad\quad \times \exp{\bigg(-\sigma_1^2
k_1^2-\sigma_2^2k_2^2-\sigma_3^2k_3^2\bigg)},
\label{CoalTransDs2317}
\end{eqnarray} }
with relative quark coordinates,

\begin{eqnarray}
&& \vec R=\frac{m_q\vec r_q+m_{\bar{q}}\vec r_{\bar{q}}
+m_{\bar{s}}\vec r_{\bar{s}}+m_c\vec r_c}{m_q+m_{\bar{q}}
+m_{\bar{s}}+m_c}, \nonumber \\
&& \vec r_1=\vec r_q-\vec r_{\bar{q}}, \nonumber \\
&& \vec r_2=\frac{m_q\vec r_q+m_{\bar{q}}\vec r_{\bar{q}}}
{m_q+m_{\bar{q}}}-\vec r_{\bar{s}}, \nonumber \\
&& \vec r_3=\frac{m_q\vec r_q+m_{\bar{q}}\vec r_{\bar{q}}
+m_{\bar{s}}\vec r_{\bar{s}}}{m_q+m_{\bar{q}}+m_{\bar{s}}}- \vec
r_c, \label{rel_coord}
\end{eqnarray}
and relative quark transverse momenta,

\begin{eqnarray}
&& \vec k=\vec p_{qT}'+\vec p_{\bar{q}T}'+\vec p_{\bar{s}T}'+\vec
p_{cT}', \nonumber \\
&& \vec k_1=\frac{m_{\bar{q}}\vec p_{qT}'-m_q\vec p_{\bar{q}T}'}
{m_q+m_{\bar{q}}}, \nonumber \\
&& \vec k_2=\frac{m_{\bar{s}}(\vec p_{qT}'+\vec p_{\bar{q}T}')-
(m_q+m_{\bar{q}})\vec p_{\bar{s}T}'}{m_q+m_{\bar{q}}+m_{\bar{s}}},
\nonumber \\
&& \vec k_3=\frac{m_c(\vec p_{qT}'+\vec p_{\bar{q}T}'+\vec
p_{\bar{s}T}')-(m_q+m_{\bar{q}}+m_{\bar{s}})\vec p_{cT}'}
{m_q+m_{\bar{q}}+m_{\bar{s}}+m_c}. \nonumber \\
\label{rel_moment}
\end{eqnarray}

In Eq. (\ref{rel_moment}), the momenta of quarks, $\vec p_i'$ are
the momenta in the rest frame of the $D_{s0}^*(2317)$ meson,
Lorentz transformed from those in the fireball frame, $\vec p_i$
as already mentioned in the previous section. The $\sigma_1$,
$\sigma_2$ and $\sigma_3$ in Eq. (\ref{CoalTransDs2317}) are
related to the oscillator frequency of the harmonic wave function
in the $s$-wave Wigner function,

\begin{eqnarray}
&& W_{D_{s0}^*(2317)}(\vec r_1, \vec r_2, \vec r_3, \vec k_1, \vec
k_2, \vec k_3) \nonumber \\
&& =8^3\exp{\bigg(-\frac{r_1^2}{\sigma_1^2}-\sigma_1^2k_1^2
\bigg)}\exp{\bigg(-\frac{r_2^2}{\sigma_2^2}-\sigma_2^2k_2^2\bigg)}
\nonumber \\
&& \times\exp{\bigg(-\frac{r_3^2}{\sigma_3^2}-\sigma_3^2
k_3^2\bigg)}, \label{wigner4}
\end{eqnarray}
via the following reduced mass, $\sigma_1^2=1/(\omega\mu_1)$,
$\sigma_2^2=1/(\omega\mu_2)$, and $\sigma_3^2=1/(\omega\mu_3)$
respectively.

\begin{eqnarray}
&& \mu_1=\frac{m_qm_{\bar{q}}}{m_q+m_{\bar{q}}}, \quad \mu_2=
\frac{(m_q+m_{\bar{q}})m_{\bar{s}}}{m_q+m_{\bar{q}}+m_{\bar{s}}},
\nonumber \\
&& \mu_3=\frac{(m_q+m_{\bar{q}}+m_{\bar{s}})m_c}{m_q+m_{\bar{q}}+
m_{\bar{s}}+m_c}.
\end{eqnarray}

In considering the relative coordinates between four quarks, two
light quarks are chosen first with the first relative coordinate
between light quarks, $\vec r_1$, then the strange quark is added
with the second relative coordinate, $\vec r_2$, and finally the
charm quark is included with the third relative coordinate, $\vec
r_3$ as shown in Eq. (\ref{rel_coord}). The four relative momenta
in Eq. (\ref{rel_moment}) follow correspondingly from the choice
of relative coordinates described above. One may wonder why the
above configuration has been chosen for the evaluation of the
transverse momentum distribution or yield of the $D_{s0}^*(2317)$
meson in a four-quark state. Actually, there are other options to
construct the relative coordinates and momenta with two light
quarks, one strange and one charm quarks in describing the
internal structure of the $D_{s0}^*(2317)$ in a four-quark state,
and thus it is possible to choose other relative coordinates and
momenta with different combination of four quarks to capture other
pictures for the $D_{s0}^*(2317)$ meson in a four-quark state.

The dependence of the transverse momentum distribution or yield on
the internal structure of a hadron has already been investigated
\cite{Cho:2019syk}, and it has been found that the choice of
relative coordinates and momenta arising from different
configurations between constituents inside the hadron does not
affect the transverse momentum distribution or yield in the
coalescence model if the Gaussian Wigner function, Eq.
(\ref{wigner4}) is applied, all quarks are in the ground states,
e.g., $s$-wave states, and only one oscillator frequency, $\omega$
is adopted, which satisfies all the conditions considered here. In
our discussion, one oscillator frequency has been determined for
all charmed hadrons, resulting in the same frequency $\omega$ for
three different $\sigma_i$'s, $\sigma_i^2=1/(\omega \mu_i)$ for
the $D_{s0}^*(2317)$ in a four-quark state. Thus, the transverse
momentum distribution of $D_{s0}^*(2317)$, Eq.
(\ref{CoalTransDs2317}), is independent of the choice of any
relative coordinates and momenta, and then configurations between
quarks inside the $D_{s0}^*(2317)$ other than Eqs.
(\ref{rel_coord}) and (\ref{rel_moment}) also leads to the same
results. Therefore, it is not possible to identify the internal
structure of the $D_{s0}^*(2317)$ from its transverse momentum
distribution by adjusting the configuration of constituent quarks
inside the $D_{s0}^*(2317)$ meson.

\begin{figure}[!t]
\begin{center}
\includegraphics[width=0.50\textwidth]{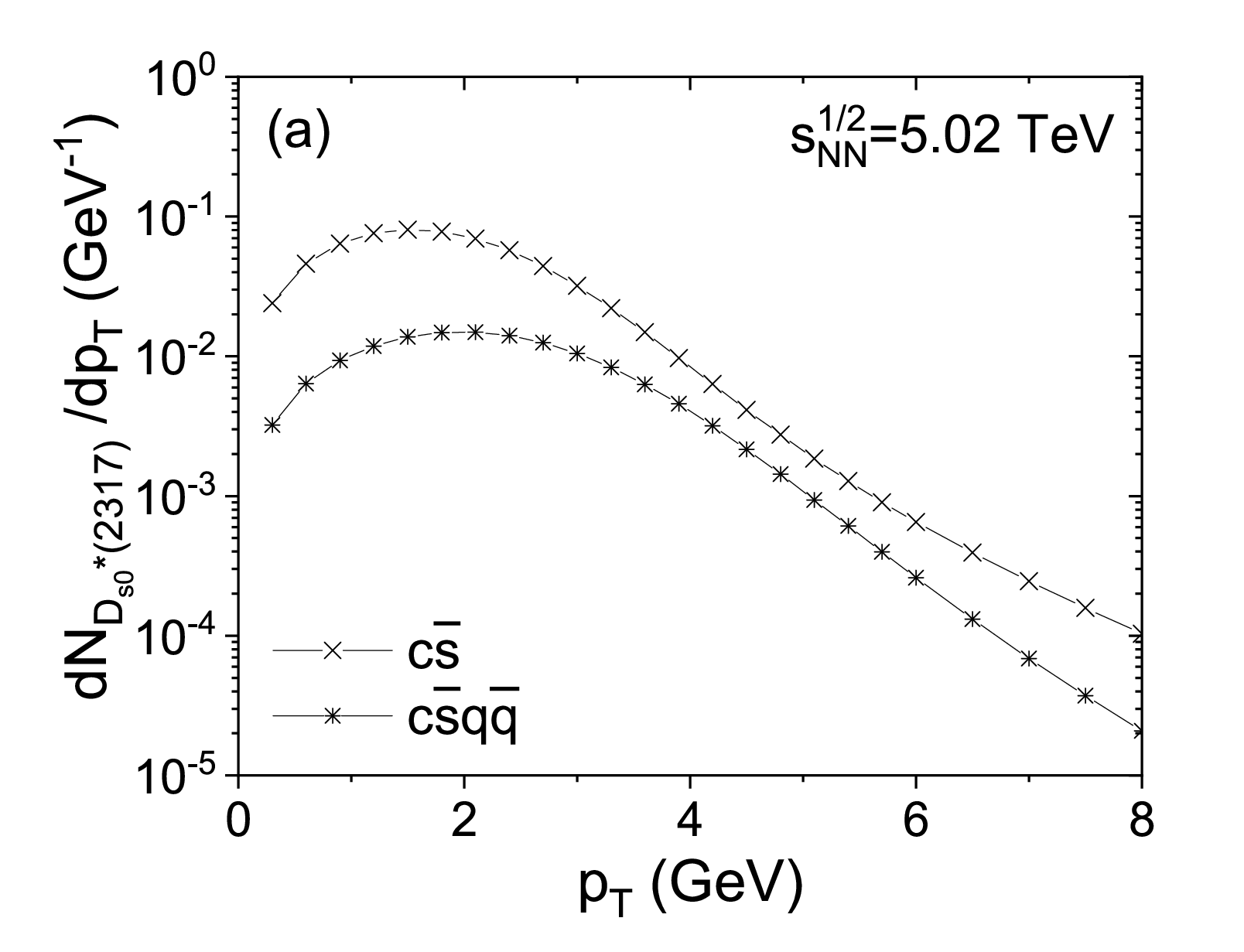}
\includegraphics[width=0.50\textwidth]{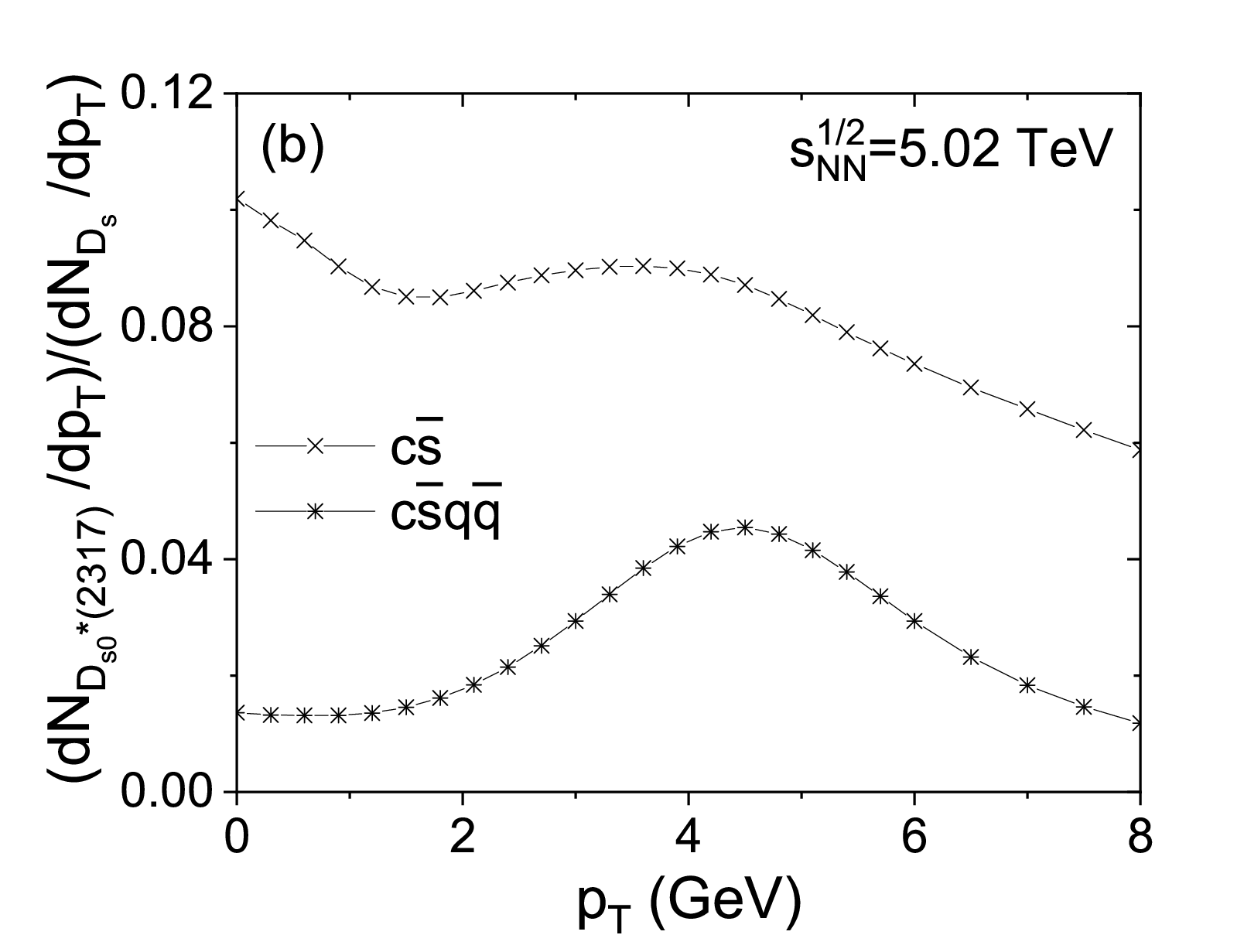}
\end{center}
\caption{(a) Transverse momentum distributions of $D_{s0}^*(2317)$
mesons in a two- and four-quark state, $dN_{D_{s0}^*(2317)}/dp_T$
at mid-rapidity in Pb+Pb collisions at $\sqrt{s_{NN}}=5.02$ TeV.
(b) Transverse momentum distribution ratios between
$D_{s0}^*(2317)$ and $D_s$, ($dN_{D_{s0}^*(2317)}/dp_T$)
/($dN_s/dp_T$) for the $D_{s0}^*(2317)$ in a two- and four-quark
state. } \label{pTdistributionDs2317q4}
\end{figure}

We calculate the transverse momentum distribution of
$D_{s0}^*(2317)$ mesons for its four-quark state using Eq.
(\ref{CoalTransDs2317}) with transverse momentum distributions of
strange, light, and charm quarks obtained in the previous
subsection. We use the same coalescence volume, 3360 fm$^3$, and
oscillator frequency, 0.103 GeV, in Eq. (\ref{CoalTransDs2317}).
As shown in Fig. \ref{pTdistributionDs2317q4}(a), the transverse
momentum distribution of the $D_{s0}^*(2317)$ in a four-quark
state is evaluated to be smaller than that of the $D_{s0}^*(2317)$
in a two-quark state. This result is consistent with one of the
conclusions on multiquark hadrons, suppressed production of a
muntiquark hadron owing to decreasing probabilities to combine
quarks to form a hadron with an increasing number of quarks within
the hadron \cite{Cho:2010db, Cho:2011ew, Cho:2017dcy}.

We also evaluate transverse momentum distribution ratios between
the $D_{s0}^*(2317)$ and $D_s$ mesons,
($dN_{D_{s0}^*(2317)}/dp_T$) /($dN_s/dp_T$) for the
$D_{s0}^*(2317)$ in a two- and four-quark state, respectively, and
show the results in Fig. \ref{pTdistributionDs2317q4}(b). The
ratio between the $D_{s0}^*(2317)$ and $D_s$ is the ratio between
mesons of similar quark contents, either $c\bar{s}/c\bar{s}$ or
$c\bar{s}q\bar{q}/c\bar{s}$, exposing the pure dependence of the
ratio on the internal structure of the $D_{s0}^*(2317)$ meson.
Moreover, as the $D_{s0}^*(2317)$ decays to $D_s \pi$ almost
entirely, one can make much clearer measurements for the ratio
between the $D_{s0}^*(2317)$ and $D_s$ meson by cancelling common
uncertainties in the analysis of both mesons.

As shown in Fig. \ref{pTdistributionDs2317q4}(b), the ratios are
much smaller than unity, reflecting both the larger mass of the
$D_{s0}^*(2317)$ compared to that of the $D_s$ and the internal
structure of the $D_{s0}^*(2317)$. The production of the
$D_{s0}^*(2317)$ in a two-quark state is suppressed owing to a
relative momentum between charm and strange quarks inside the
$D_{s0}^*(2317)$, a $p$-wave, thereby yielding to its reduced
production compared to a hadron with a $s$-wave, i.e., the $D_s$
meson. On the other hand, the production of the $D_{s0}^*(2317)$
in a four-quark state is much more suppressed due to much smaller
chances for four quarks to coalesce to form the $D_{s0}^*(2317)$.

Nevertheless, the two ratios differ maximally in magnitude by a
factor of six or seven at low transverse momentum regions. The
transverse momentum distribution ratio between the
$D_{s0}^*(2317)$ meson in a four-quark state and $D_s$ meson
varies from 0.013 at $p_T=0$ GeV up to 0.045, showing a peak at
about $p_T=4.5$ GeV. It is interesting to note that the peak in
the transverse momentum distribution ratio with the
$D_{s0}^*(2317)$ in a four-quark state resembles the transverse
momentum distribution ratio between a baryon and a meson, i.e., an
anti-proton and a pion \cite{Greco:2003mm, Greco:2003xt}. Two
extra light quarks, occupying approximately a fifth of the total
momentum of the $D_{s0}^*(2317)$ in a four-quark state, play
similar roles as one quark in the baryon, enhancing the transverse
momentum distribution ratio at the intermediate transverse
momentum regions as a result of the overlap between exponential-
and power-type functions in the light quark transverse momentum
distributions when the $D_{s0}^*(2317)$ in a four-quark state is
formed.

Carrying out the integration of the transverse momentum
distribution of the $D_{s0}^*(2317)$ meson in a four-quark state
over the entire transverse momentum region results in 0.0419 for
its yield, smaller by about a factor of five compared to the yield
of the $D_{s0}^*(2317)$ in a two-quark state, 0.192 as shown in
the Table \ref{charmestrangeyields}.

\begin{table}[!t]
\caption{Yields of the $D_{s0}^*(2317)$ for two- and four-quark
states at mid-rapidity evaluated from the integration of the
transverse momentum distributions shown in Figs.
\ref{pTdistributionDs2317q4} over all transverse momenta, together
with previous results for the yields of the $D_{s0}^*(2317)$ in
two-quark, four-quark, and hadronic molecule states expected at
$\sqrt{s_{NN}}=5.02$ TeV at LHC \cite{Cho:2017dcy}. }
\label{Ds2317yields}
\begin{center}
\begin{tabular}{c|c|c|c|c}
\hline \hline
Yields & Ther. & $c\bar{s}$ & $c\bar{s}q\bar{q}$ & Mol.  \\
\hline Coal. & ~0.180~ & ~0.192~ & ~0.0419~ &   \\
ExHIC \cite{Cho:2017dcy} & ~0.19~ & ~0.064~ & ~0.0057~ & ~0.018~ \\
\hline \hline
\end{tabular}
\end{center}
\end{table}

We summarize the yield of the $D_{s0}^*(2317)$ meson in both two-
and four-quark states evaluated here in Table \ref{Ds2317yields},
together with previous evaluations for the yields of the
$D_{s0}^*(2317)$ in two-quark, four-quark, and hadronic molecule
states expected at $\sqrt{s_{NN}}=5.02$ TeV at LHC
\cite{Cho:2017dcy}. The results are all larger compared to the
previous expectations, mainly originating from the charm quark
numbers evaluated here, $N_c=17.4$, which is larger than before,
$N_c=14$ \cite{Cho:2017dcy}. Nevertheless, the difference of the
$D_{s0}^*(2317)$ yields for two- and four-quark states is still
large enough in both calculations, enabling to identify the
internal structure of the $D_{s0}^*(2317)$ from its production in
relativistic heavy ion collisions.

\section{Conclusions}

We have studied charm-strange mesons, i.e., the $D_s$, $D_s^*$,
$D_{s0}^*(2317)$, and $D_{s1}(2460)$ by focusing on their
production from a quark-gluon plasma at $\sqrt{s_{NN}}=5.02$ TeV
in relativistic heavy ion collisions. We have investigated the
transverse momentum distributions of $\phi$ and $D^0$ mesons
measured by ALICE Collaboration at $\sqrt{s_{NN}}=5.02$ TeV at LHC
in order to find information on the transverse momentum
distributions of strange and charm quarks at $\sqrt{s_{NN}}=5.02$.
Then, we have calculated the transverse momentum distributions and
yields of $D_s$, $D_s^*$, $D_{s0}^*(2317)$, and $D_{s1}(2460)$
mesons based on the coalescence model.

We have compared the transverse momentum distribution of the $D_s$
evaluated in this work to that of the $D_s$ measured at LHC by
ALICE Collaboration. Moreover, we have assessed the transverse
momentum ratio between $D_s$ and $D^0$ mesons in order to
investigate the role of light and strange quarks in the
coalescence production of charmed mesons in heavy ion collisions.
Furthermore, we have calculated the transverse momentum
distribution of the $D_{s0}^*(2317)$ meson in a four-quark state,
and have compared to that of the $D_{s0}^*(2317)$ meson in a
two-quark state to find a way of differentiating the quark
structure of the $D_{s0}^*(2317)$ meson in relativistic heavy ion
collision experiments.

The transverse momentum distribution of the $D_s$ meson evaluated
with new transverse momentum distributions of charm and strange
quarks at $\sqrt{s_{NN}}=5.02$ TeV obtained in this work is found
to agree reasonably well with the experimental measurement by
ALICE Collaboration. In addition, the yield of the $D_s$ meson
obtained by integrating the transverse momentum distribution of
$D_s$ meson over all transverse momentum regions is observed to be
close to the measurement within about $10\%$.

We find that the transverse momentum distribution ratio between
$D_s$ and $D^0$ mesons reflects the different interplay between
constituents when they are produced by coalescence in heavy ion
collisions. The ratio is found to be larger with a peak at low
$p_T$ regions than at high $p_T$ regions, implying the enhanced
production of $D_s$ mesons compared to $D^0$ mesons at low $p_T$
regions due to the strangeness enhancement in heavy ion
collisions. On the other hand, the ratio is shown to decrease with
increasing transverse momenta as contributions to $D^0$ mesons
from charm quarks by fragmentation prevail at high $p_T$ regions,
about 60.86$\%$, Eq. (\ref{feeddownD0}) much larger than those to
$D_s$ mesons, about 8.02$\%$, Eq. (\ref{feeddown}).

The yield, or the transverse momentum distribution of the
$D_{s0}^*(2317)$ in a four-quark state is evaluated to be smaller
than that of the $D_{s0}^*(2317)$ in a two-quark state, as has
been expected. The result reflects the suppressed production for a
hadron with more quarks, as the probability to combine more quarks
to form the hadron decreases with increasing number of quarks
within the hadron. Therefore, it is at least possible to
discriminate the quark contents of the $D_{s0}^*(2317)$, whether
it exists in a compact two- or four-quark states from the
measurement of the yield and transverse momentum distribution of
$D_{s0}^*(2317)$ in heavy ion collisions. We hope that we identify
the quark structure of the $D_{s0}^*(2317)$ meson by comparing the
yield and transverse momentum distribution of the $D_{s0}^*(2317)$
obtained here to that measured in relativistic heavy ion collision
experiments in the near future.

% In order to understand in more detail the structure of the
% $D_{s0}^*(2317)$ from its production in relativistic heavy ion
% collisions, it is also necessary to investigate the hadronic
% effects on the $D_{s0}^*(2317)$ during the hadronic stages as have
% done in the $X(3872)$ meson \cite{Cho:2013rpa, Torres:2014fxa}.

As shown in Fig. \ref{pTdistribution_charmstrange} and Table.
\ref{charmestrangeyields}, the feed-down contributions from
$D_s^*$, $D_{s0}^*(2317)$, and $D_{s1}(2460)$ mesons to the $D_s$
are responsible for more than $75.0\%$ of the total $D_s$ meson
production. Thus, it must be essential to understand the
production of the above charm-strange mesons experimentally in
order to explain in detail the production of $D_s$ meson. We hope
that the yields and transverse momentum distributions of $D_s^*$,
$D_{s0}^*(2317)$, and $D_{s1}(2460)$ mesons presented here would
be helpful in measuring those mesons in future heavy ion collision
experiments.

Therefore, studying the production of charm-strange mesons in
relativistic heavy ion collisions provides us with opportunities
for a better understanding of the hadronization processes
involving charm and strange quarks in the quark-gluon plasma. We
hope that transverse momentum distributions of light, strange, and
charm quarks as well as those of the $D_s$, $D_s^*$,
$D_{s0}^*(2317)$, and $D_{s1}(2460)$ meson evaluated here would
help understand various kinds of hadrons produced in heavy ion
collisions, thereby contributing to revealing the properties of
the quark-gluon plasma.

\section*{Acknowledgements}

This work was supported by the National Research Foundation of
Korea (NRF) grant funded by the Korea government (MSIT) No.
RS-2023-00280831 (S.C.), No. 2023R1A2C300302311, and No.
2023K2A9A1A0609492411 (S.H.L.).

\end{document}